\documentclass[graybox]{svmult}

\usepackage{mathptmx}       
\usepackage{helvet}         
\usepackage{courier}        
\usepackage{type1cm}        
\usepackage{makeidx}         
\usepackage{graphicx}        
\usepackage{multicol}        
\usepackage[bottom]{footmisc}

\newif\ifcomments 

\usepackage{graphicx}
\graphicspath{ {Figures/} }

\usepackage[hyphens]{url}

\makeindex             

\begin{document}


\title*{Serverless Computing: Current Trends and Open
Problems}


\author{Ioana Baldini, Paul Castro, Kerry Chang, Perry Cheng, Stephen Fink,
   Vatche Ishakian, Nick Mitchell, Vinod Muthusamy, Rodric Rabbah,
   Aleksander Slominski, Philippe Suter
}


\institute{Ioana Baldini, Paul Castro, Kerry Chang, Perry Cheng, Stephen Fink,
   Nick Mitchell, Vinod Muthusamy, Rodric Rabbah,
   Aleksander Slominski, Philippe Suter \at IBM Research, \email{name@us.ibm.com}
\and Vatche Ishakian \at Bentley University  \email{vishakian@bentley.edu}}





%
%
\authorrunning{ }
\maketitle

\abstract{Serverless computing has emerged as a new compelling
paradigm for the deployment of applications and services. It
represents an evolution of cloud programming models, abstractions, and
platforms, and is a testament to the maturity and wide adoption of
cloud technologies. In this chapter, we survey existing serverless
platforms from industry, academia, and open source projects, identify
key characteristics and use cases, and describe technical challenges
and open problems.}







\section{Introduction}

Serverless Computing (or simply serverless) is emerging as a new and
compelling paradigm for the deployment of cloud applications, largely
due to the recent shift of enterprise application architectures to
containers and microservices~\cite{microserviceadoption}.
Figure~\ref{fig:serverless-term} below shows the increasing popularity
of the ``serverless" search term over the last five years as reported
by Google Trends. This is an indication of the increasing attention
that serverless computing has garnered in industry tradeshows,
meetups, blogs, and the development community. By contrast, the
attention in the academic community has been limited.

\begin{figure}[h]
\includegraphics[width=\textwidth]{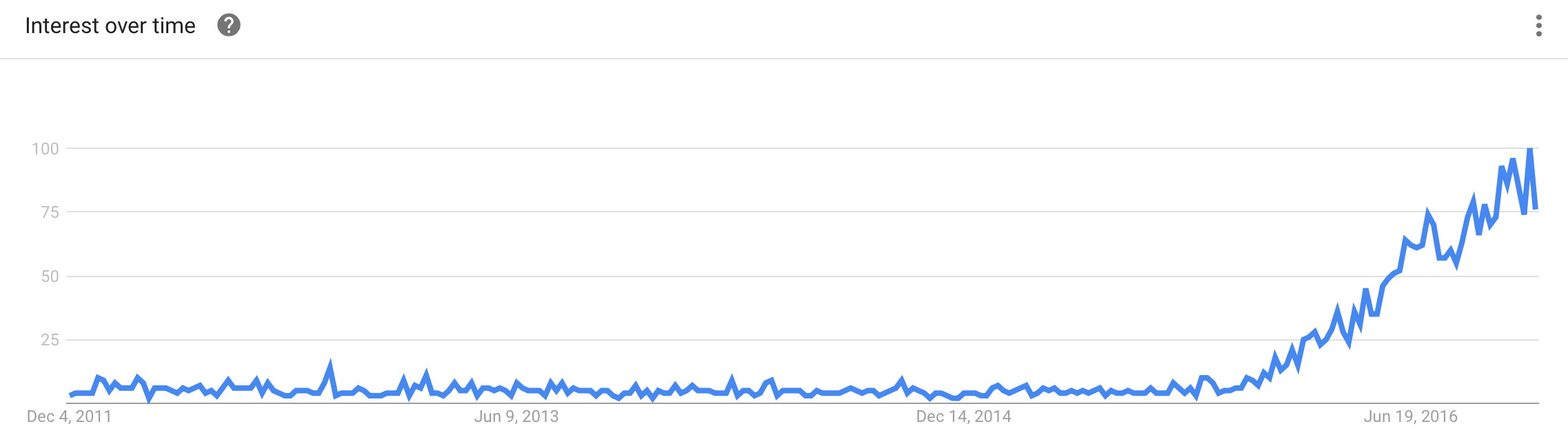}
\caption{Popularity of the term ``serverless" as reported by Google
Trends}
\label{fig:serverless-term}
\end{figure}

From the perspective of an Infrastructure-as-a-Service (IaaS)
customer, this paradigm shift presents both an opportunity and a risk.
On the one hand, it provides developers with a simplified programming
model for creating cloud applications that abstracts away most, if not
all, operational concerns; it lowers the cost of deploying cloud code
by charging for execution time rather than resource allocation; and it
is a platform for rapidly deploying small pieces of cloud-native code
that responds to events, for instance, to coordinate microservice
compositions that would otherwise run on the client or on dedicated
middleware.  On the other hand, deploying such applications in a
serverless platform is challenging and requires relinquishing to the
platform design decisions that concern, among other things,
quality-of-service (QoS) monitoring, scaling, and fault-tolerance
properties.

From the perspective of a cloud provider, serverless computing
provides an additional opportunity to control the entire development
stack, reduce operational costs by efficient optimization and
management of cloud resources, offer a platform that encourages the
use of additional services in their ecosystem, and lower the effort
required to author and manage cloud-scale applications.

Serverless computing is a term coined by industry to describe a
programming model and architecture where small code snippets are
executed in the cloud without any control over the resources on which
the code runs. It is by no means an indication that there are no
servers, simply that the developer should leave most operational
concerns such as resource provisioning, monitoring, maintenance,
scalability, and fault-tolerance to the cloud provider.

The astute reader may ask how this differs from the
Platform-as-a-Service (PaaS) model, which also abstracts away the
management of servers. A serverless model provides a ``stripped down"
programming model based on stateless functions.
Unlike PaaS, developers can write arbitrary code and are not limited
to using a pre-packaged application.  The version of serverless that
explicitly uses functions as the deployment unit is also called
Function-as-a-Service (FaaS).

Serverless platforms promise new capabilities that make writing
scalable microservices easier and cost effective, positioning
themselves as the next step in the evolution of cloud computing
architectures. Most of the prominent cloud computing providers
including Amazon \cite{aws-lambda}, IBM \cite{ibm-openwhisk},
Microsoft \cite{microsoft-azure-functions}, and Google
\cite{google-cloud-functions} have recently released serverless
computing capabilities. There are also several open-source efforts
including the OpenLambda project~\cite{open-lambda}.

Serverless computing is in its infancy and the research community has
produced only a few citable publications at this time. OpenLambda
\cite{open-lambda} proposes a reference architecture for serverless
platforms and describes challenges in this space (see Section
\ref{architecture-upcoming}) and we have previously published two of
our use-cases ~\cite{Baldini2016, Mengting2016} (see Section
\ref{use-cases-event-processing}). There are also several books for
practitioners that target developers interested in building
applications using serverless platforms~\cite{FernandezBook,
ManningBook}.


\subsection{Defining Serverless}

\begin{figure}[h]
\includegraphics[width=\textwidth]{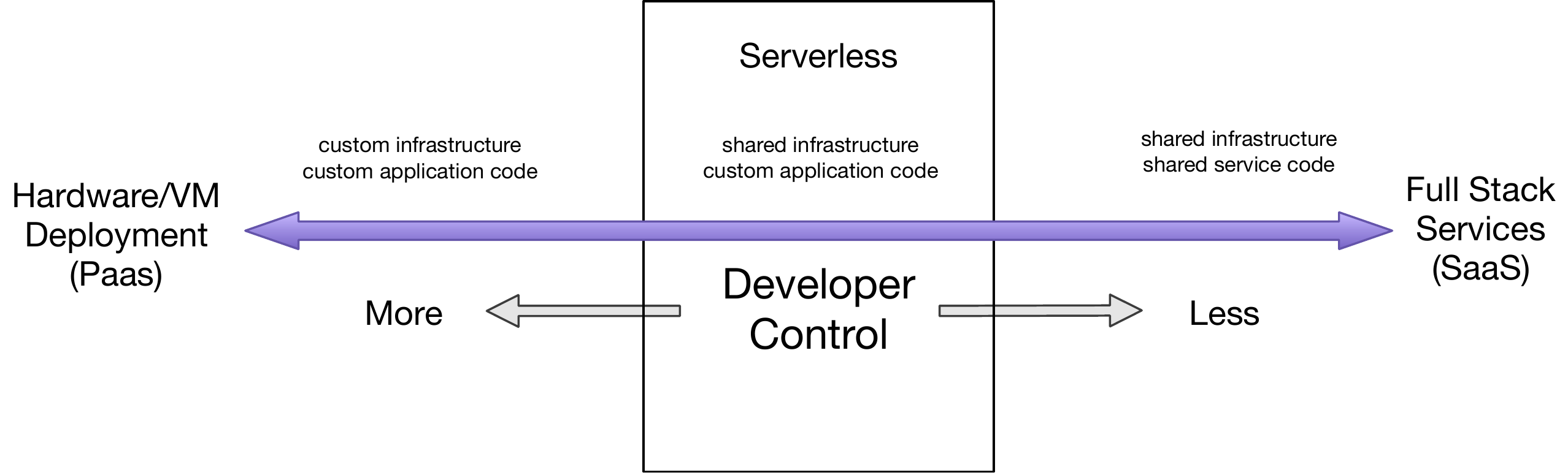}
\caption{Developer control and Serverless computing}
\label{fig:position}
\end{figure}

Succinctly defining the term serverless can be difficult as the
definition will overlap with other terms such as PaaS and
Software-as-a-Service (SaaS). One way to explain serverless is to
consider the varying levels of developer control over the cloud
infrastructure, as illustrated in Figure~\ref{fig:position}. The
Infrastructure-as-a-Service (IaaS) model is where the developer has
the most control over both the application code and operating
infrastructure in the cloud. Here, the developer is responsible for
provisioning the hardware or virtual machines, and can customize every
aspect of how an application gets deployed and executed.  On the
opposite extreme are the PaaS and SaaS models, where the developer is
unaware of any infrastructure, and consequently no longer has control
over the infrastructure.  Instead, the developer has access to
pre-packaged components or full applications.  The developer is
allowed to host code here, though that code may be tightly coupled to
the platform.

For this chapter, we will focus on the space in the middle of
Figure~\ref{fig:position}. Here, the developer has control over the
code they deploy into the Cloud, though that code has to be written in
the form of stateless functions. (The reason for this will be
explained in Section~\ref{sec:architecture}.) The developer does not
worry about the operational aspects of deployment and maintenance of
that code and expects it to be fault-tolerant and auto-scaling.  In
particular, the code may be scaled to zero where no servers are
actually running when the user's function code is not used, and there
is no cost to the user. This is in contrast to PaaS solutions where
the user is often charged even during idle periods.

There are numerous serverless platforms the fall into the above
definition.  In this chapter, we present the architecture and other
relevant features of serverless computing, such as the programming
model.  We also identify the types of application workloads that are
suitable to run on serverless computing platforms. We then conclude
with open research problems, and future research challenges. Many of these
challenges are a pressing need in industry and could benefit from
contributions from academia.

\section{Evolution}

Serverless computing was popularized by Amazon in the re:Invent 2014
session ``Getting Started with AWS Lambda"~\cite{aws-lambda-2014}.
Other vendors followed in 2016 with the introduction of Google Cloud
Functions~\cite{google-cloud-functions}, Microsoft Azure
Functions~\cite{microsoft-azure-functions} and IBM
OpenWhisk~\cite{ibm-openwhisk}. However, the serverless approach to
computing is not completely new. It has emerged following recent
advancements and adoption of virtual machine (VM) and then container
technologies.  Each step up the abstraction layers led to more
lightweight units of computation in terms of resource consumption,
cost, and speed of development and deployment.


Among existing approaches, Mobile Backend as-a-Service (MBaaS) bears a
close resemblance to serverless computing.
Some of those services even provided ``cloud functions", that is, the
ability to run some code server-side on behalf of a mobile app without
the need to manage the servers.  An example of such a service is
Facebook's Parse Cloud Code \cite{parse-cloud-code}. Such code,
however, was typically limited to mobile use cases.

Software-as-a-Service (SaaS) may support the server-side execution of
user provided functions but they are executing in the context of an
application and hence limited to the application domain. Some SaaS vendors allow the integration of arbitrary code hosted somewhere else and invoked via an API call.  For example, this is approach is used by the Google Apps Marketplace in Google Apps for Work \cite{google-apps-marketplace}.



\section{Architecture} \label{sec:architecture}

There are a lot of misconceptions surrounding serverless starting with
the name. Servers are still needed, but developers need not concern
themselves with managing those servers. Decisions such as the number
of servers and their capacity are taken care of by the serverless
platform, with server capacity automatically provisioned as needed by
the workload. This provides an abstraction where computation (in the
form of a stateless function) is disconnected from where it is going
to run.

\begin{figure}[h]
\includegraphics[width=\textwidth]{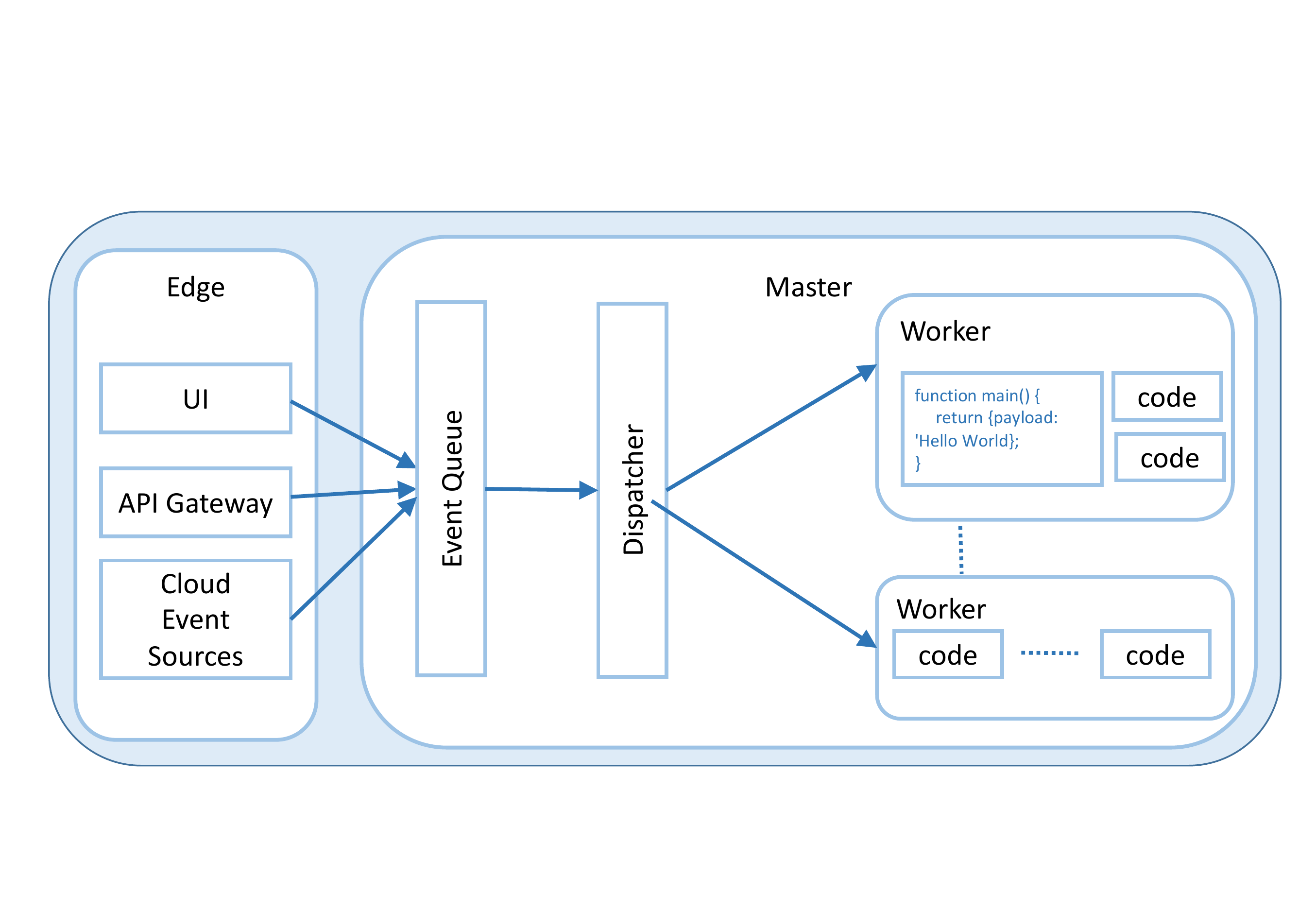}
\caption{Serverless platform architecture
}
\label{fig:architecture}
\end{figure}

The core capability of a serverless platform is that of an event
processing system, as depicted in Figure~\ref{fig:architecture}.  The
service must manage a set of user defined functions, take an event
sent over HTTP or received from an event source, determine which
function(s) to which to dispatch the event, find an existing instance
of the function or create a new instance, send the event to the
function instance, wait for a response, gather execution logs, make
the response available to the user, and stop the function when it is
no longer needed.

The challenge is to implement such functionality while considering
metrics such as cost, scalability, and fault tolerance. The platform
must quickly and efficiently start a function and process its input.
The platform also needs to queue events, and based on the state of the
queues and arrival rate of events, schedule the execution of
functions, and manage stopping and deallocating resources for idle
function instances. In addition, the platform needs to carefully
consider how to scale and manage failures in a cloud environment.

\subsection{Survey of serverless platforms}


In this section we will compare a number of serverless platform.  We
first list the dimensions which will be used to characterize the
architectures of these platforms, followed by a brief description of
each platform.

\subsubsection{Characteristics}

There are a number of characteristics that help distinguish the
various serverless platforms. Developers should be aware of these
properties when choosing a platform.

\begin{itemize}
\item \emph{Cost}: Typically the usage is metered and users pay only for the time and resources used when serverless functions are running.  This ability to scale to zero instances is one of the key differentiators of a serverless platform. The resources that are metered, such as memory or CPU, and the pricing model, such as off-peak discounts, vary among providers.
\item \emph{Performance and limits}: There are a variety of limits set on the runtime resource requirements of serverless code, including the number of concurrent requests, and the maximum memory and CPU resources available to a function invocation. Some limits may be increased when users' need grow, such as the concurrent request threshold, while others are inherent to the platforms, such as the maximum memory size.
\item \emph{Programming languages}: Serverless services support a wide variety of programming languages including Javascript, Java, Python, Go, C\#,  and Swift. Most platforms support more than one programming language. Some of the platforms also support extensibility mechanisms for code written in any language as long as it is packaged in a Docker image that supports a well-defined API.
\item \emph{Programming model}: Currently, serverless platforms typically execute a single \texttt{main} function that takes a dictionary (such as a JSON object) as input and produces a dictionary as output.
\item \emph{Composability}: The platforms generally offer some way to invoke one serverless function from another, but some platforms provide higher level mechanisms for composing these functions and may make it easier to construct more complex serverless apps.
\item \emph{Deployment}: Platforms strive to make deployment as simple
as possible. Typically, developers just need to provide a file with the function source code. Beyond that there are many options where code can be packaged as an archive with multiple files inside or as a Docker image with binary code. As well, facilities to version or group functions are useful but rare.
\item \emph{Security and accounting}: Serverless platforms are multi-tenant and must isolate the execution of functions between users and provide detailed accounting so users understand how much they need to pay.
\item \emph{Monitoring and debugging}: Every platform supports basic debugging by using print statements that are recorded in the execution logs.  Additional capabilities may be provided to help developers find bottlenecks, trace errors, and better understand the cicumstances of function execution.
\end{itemize}


\subsubsection{Commercial platforms}

Amazon's AWS Lambda~\cite{aws-lambda} was the first serverless
platform and it defined several key dimensions including cost,
programming model, deployment, resource limits, security, and
monitoring.  Supported languages include Node.js, Java, Python, and
C\#.  Initial versions had limited composability but this has been
addressed recently. The platform takes advantage of a large AWS
ecosystem of services and makes it easy to use Lambda functions as
event handlers and to provide glue code when composing services.

Currently available as an Alpha release, Google Cloud
Functions~\cite{google-cloud-functions} provides basic FaaS
functionality to run serverless functions written in Node.js in
response to HTTP calls or events from some Google Cloud services.
The functionality is currently limited but expected to grow in future
versions.

Microsoft Azure Functions~\cite{microsoft-azure-functions} provides
HTTP webhooks and integration with Azure services to run user provided
functions. The platform supports C\#, F\#, Node.js, Python, PHP,
bash, or any executable.
The runtime code is
open-source and available on GitHub under an MIT License.  To ease
debugging, the Azure Functions CLI provides a local development
experience for creating, developing, testing, running, and debugging
Azure Functions.

\begin{figure}[h]
\includegraphics[width=\textwidth]{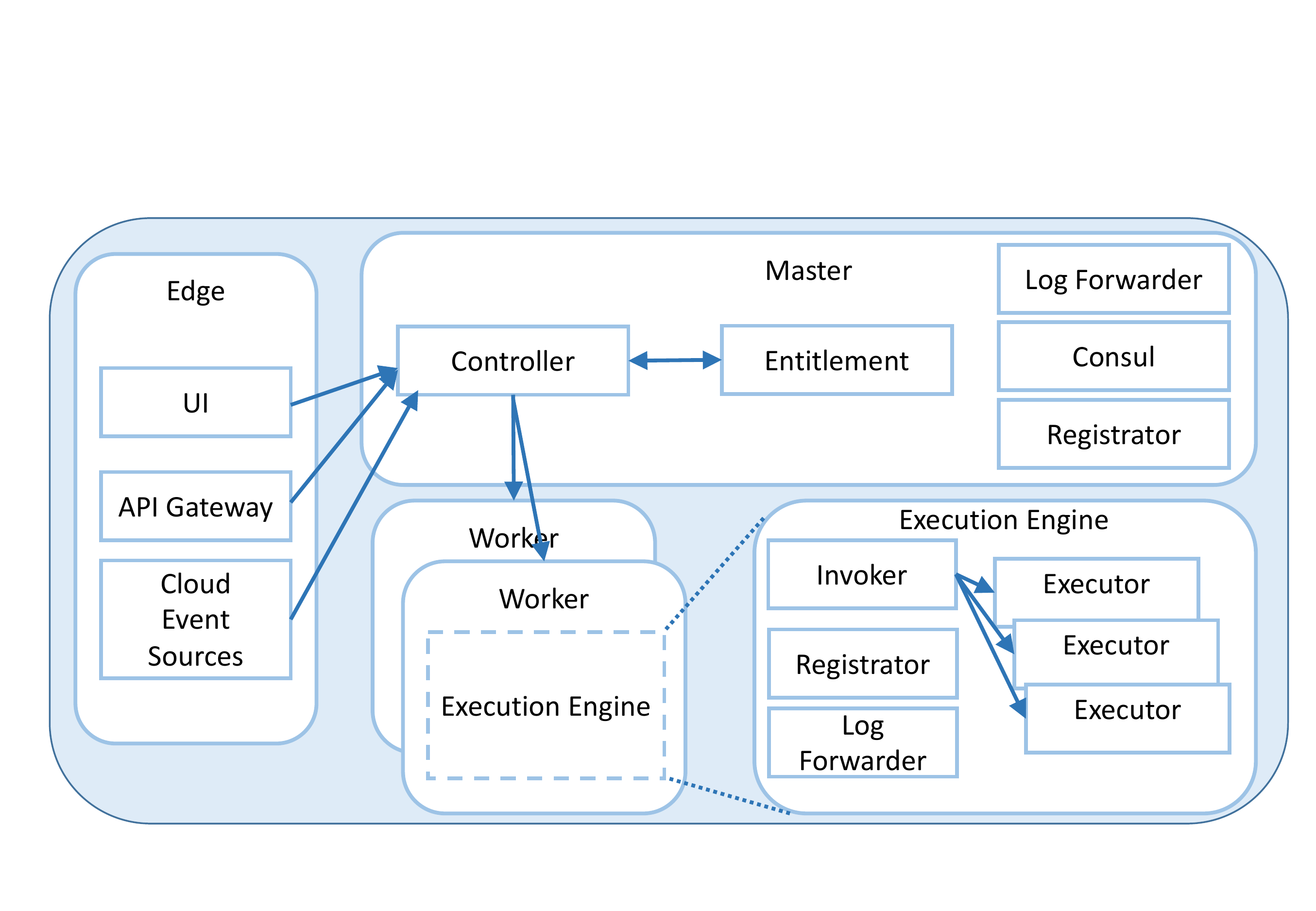}
\caption{IBM OpenWhisk architecture}
\label{fig:openwhisk}
\end{figure}

IBM OpenWhisk~\cite{ibm-openwhisk} provides event-based serverless
programming with the ability to chain serverless functions to create
composite functions.  It supports Node.js, Java, Swift, Python, as well as arbitrary binaries embedded in a Docker container. OpenWhisk is available on GitHub
under an Apache open source license. The main architectural components
of the OpenWhisk platform are shown in Figure~\ref{fig:openwhisk}.
Compared to the generic architectural diagram in
Figure~\ref{fig:architecture}, we can see there are additional
components handling important requirements such as security,
logging, and monitoring.

%
%
%
%
%
%
%
%
%
%

\subsubsection{New and upcoming serverless platforms}
\label{architecture-upcoming}

There are several serverless projects ranging from open source
projects to vendors that find serverless a natural fit
for their business.

OpenLambda~\cite{open-lambda} is an open-source serverless computing
platform.  The source code is available in GitHub under an Apache
License. The OpenLambda paper~\cite{HendricksonSHVA16} outlines a
number of challenges around performance such as supporting faster
function startup time for heterogenous language runtimes and across a
load balanced pool of servers, deployment of large amounts of code,
supporting stateful interactions (such as HTTP sessions) on top of
stateless functions, using serverless functions with databases and
data aggregators, legacy decomposition, and cost debugging. We have
identified similar challenges in Section~\ref{sec:challenges}.

Some serverless systems are created by companies that see the need for
serverless computing in the environments they operate. For example
Galactic Fog~\cite{galacticfog} added serverless computing to their
Gestalt Framework running on top of Mesos D/C.  The source code is
available under an Apache 2 license.
Auth0 has created webtasks~\cite{auth0-webtask} that execute serverless functions
to support webhook endpoints used in complex security scenarios. This code also available as open source.
Iron.io had a serverless support for tasks since 2012~\cite{iron-io-webtask}.
Recently they announced Project Kratos~\cite{iron-io-lambda} that allows developers
to convert AWS Lambda functions into Docker images, and is available
under an Apache 2 license.
Additionally they are working with Cloud Foundry to bring multi-cloud
serverless support to Cloud Foundry users~\cite{iron-io-cf}.
LeverOS is an open source project that uses an RPC model to communicate between services. Computing resources in LeverOS can be tagged so repeated function invocations can be targeted to a specific container to optimize runtime performance, such as taking advantage of warm caches in a container.\cite{leveros}.

\subsection{Benefits and drawbacks}

Compared to IaaS platforms, serverless architectures offer different
tradeoffs in terms of control, cost, and flexibility. In particular,
they force application developers to carefully think about the cost of
their code when modularizing their applications, rather than latency,
scalability, and elasticity, which is where significant development
effort has traditionally been spent.

The serverless paradigm has advantages for both consumers and
providers.  From the consumer perspective, a cloud developer no longer
needs to provision and manage servers, VMs, or containers as the basic
computational building block for offering distributed services.
Instead the focus is on the business logic, by defining a set of
functions whose composition enables the desired application behavior.
The stateless programming model gives the provider more control
over the software stack, allowing them to,
among other things, more transparently deliver security patches and
optimize the platform.

There are, however, drawbacks to both consumers and providers. For
consumers, the FaaS model offered by the platform may be too
constraining for some applications. For example, the platform may not
support the latest Python version, or certain libraries may not be
available. For the provider, there is now a need to manage issues such
as the lifecycle of the user's functions, scalability, and fault
tolerance in an application-agnostic manner. This also means that
developers have to carefully understand how the platform behaves and
design the application around these capabilities.

One property of serverless platforms that may not be evident at the
outset is that the provider tends to offer an ecosystem of services
that augment the user's functions. For example, there may be services
to manage state, record and monitor logs, send alerts, trigger events,
or perform authentication and authorization. Such rich ecosystems can
be attractive to developers, and present another revenue opportunity
for the cloud provider.  However, the use of such services brings with
it a dependence on the provider's ecosystem, and a risk of vendor
lock-in.

\subsection{Current state of serverless platforms}

There are many commonalities between serverless platforms. They share
similar pricing, deployment, and programming models. The main
difference among them is the cloud ecosystem: current serverless
platforms only make it easy to use the services in their own ecosystem
and the choice of platform will likely force developers to use the
services native to that platform.  That may be changing as open source
solutions may work well across multiple cloud platforms.

\section{Programming model}

Serverless functions have limited expressiveness as they are built to
scale. Their composition may be also limited and tailored to support
cloud elasticity. To maximize scaling, serverless functions do not
maintain state between executions. Instead, the developer can write
code in the function to retrieve and update any needed state.  The
function is also able to access a context object that represents the
environment in which the function is running (such as a security
context). For example, a function written in JavaScript could take the
input, as a JSON object, as the first parameter, and context as the
second:

\begin{verbatim}
function main(params, context) {
   return {payload:  'Hello, ' + params.name
                  + ' from ' + params.place};
}
\end{verbatim}


\subsection{Ecosystem}

Due to the limited and stateless nature of serverless functions, an
ecosystem of scalable services that support the different
functionalities a developer may require is essential to having a
successfully deployed serverless application. For example, many applications will require the serverless function to retrieve state from permanent storage  (such as a
file server or database).  There may be an existing ecosystem of functions that support API calls to various storage systems.  While the functions themselves may scale due to the serverless guarantees, the underlying storage system itself must provide reliability and QoS guarantees to ensure smooth operation.  Serverless functions can be used to coordinate any number of systems such as identity providers,
messaging queues, and cloud-based storage. Dealing with the challenges of
scaling of these systems on-demand is as critical but outside the control of the serverless platform. To increase the adoption of
serverless computing there is a need to provide such scalable
services. Such an ecosystem enables ease of integration and fast
deployment at the expense of vendor lock-in.

\subsection{Tools and frameworks}

Creating and managing serverless functions requires several
operations. Instead of managing each function independently it is much
more convenient to have a framework that can logically group functions
together to deploy and update them as a unit. A framework may
also make it easier to create functions that are not bound to one
serverless service provider by providing abstractions that hide
low-level details of each serverless provider. Other frameworks may take
existing popular programming models and adapt them for serverless
execution. For example Zappa \cite{zappa} and Chalice \cite{chalice}
use an @app.route decorator to make it possible to write python code
that looks like a webserver but can be deployed as a serverless
function:

\begin{verbatim}
@app.route("/{name}/{place}")
def index():
    return {"hello": name, "from": place }
\end{verbatim}


\section{Use cases and workloads}

Serverless computing has been utilized to support a wider range of applications.
From a functionality perspective, serverless and more traditional
architectures may be used interchangeably. The determination of when
to use serverless will likely be influenced by other non-functional
requirements such as the amount of control over operations required,
cost, as well as application workload characteristics.

From a cost perspective, the benefits of a serverless architecture are
most apparent for \emph{bursty, compute intensive} workloads. Bursty
workloads fare well because the developer offloads the elasticity of
the function to the platform, and just as important, the function can
scale to zero, so there is no cost to the consumer when the system is
idle. Compute intensive workloads are appropriate since in most
platforms today, the price of a function invocation is proportional to
the running time of the function. Hence, I/O bound functions are
paying for compute resources that they are not fully taking advantage
of. In this case, a multi-tenant server application that multiplexes
requests may be cheaper to operate.

From a programming model perspective, the stateless nature of
serverless functions lends themselves to application structure similar
to those found in functional reactive programming~\cite{frp}. This
includes applications that exhibit event-driven and flow-like
processing patterns.

\subsection{Event processing}
\label{use-cases-event-processing}

\begin{figure}[h]
\includegraphics[width=\textwidth]{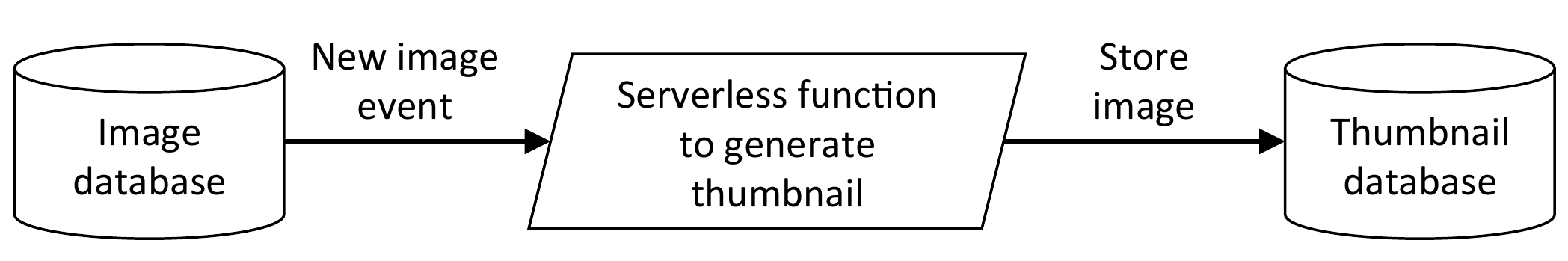}
\caption{Image processing}
\label{fig:UseCase-Thumbnail}
\end{figure}

One class of applications that are very much suitable for is
event-based programming~\cite{Baldini2016,Mengting2016}. The most basic example,
popularized by AWS Lambda, that has become the ``Hello World" of
serverless computing is a simple image processing event handler
function. The function is connected to a data store, such as Amazon
S3~\cite{s3}, that emits change events. Each time a new image file is
uploaded to a folder in S3 an event is generated, and forwarded to the
event handler function that generates a thumbnail image that is stored
in another S3 folder. The flow is depicted in
Figure~\ref{fig:UseCase-Thumbnail}. This example works well for
serverless demos as the function is completely stateless and
idempotent which has the advantage that in the case of failure (such
as network problems accessing the S3 folder), the function can be
executed again with no side effects. It is also an exemplary use case
of a bursty, compute intensive workload as described above.

\subsection{API composition}

\begin{figure}[h]
\includegraphics[width=\textwidth]{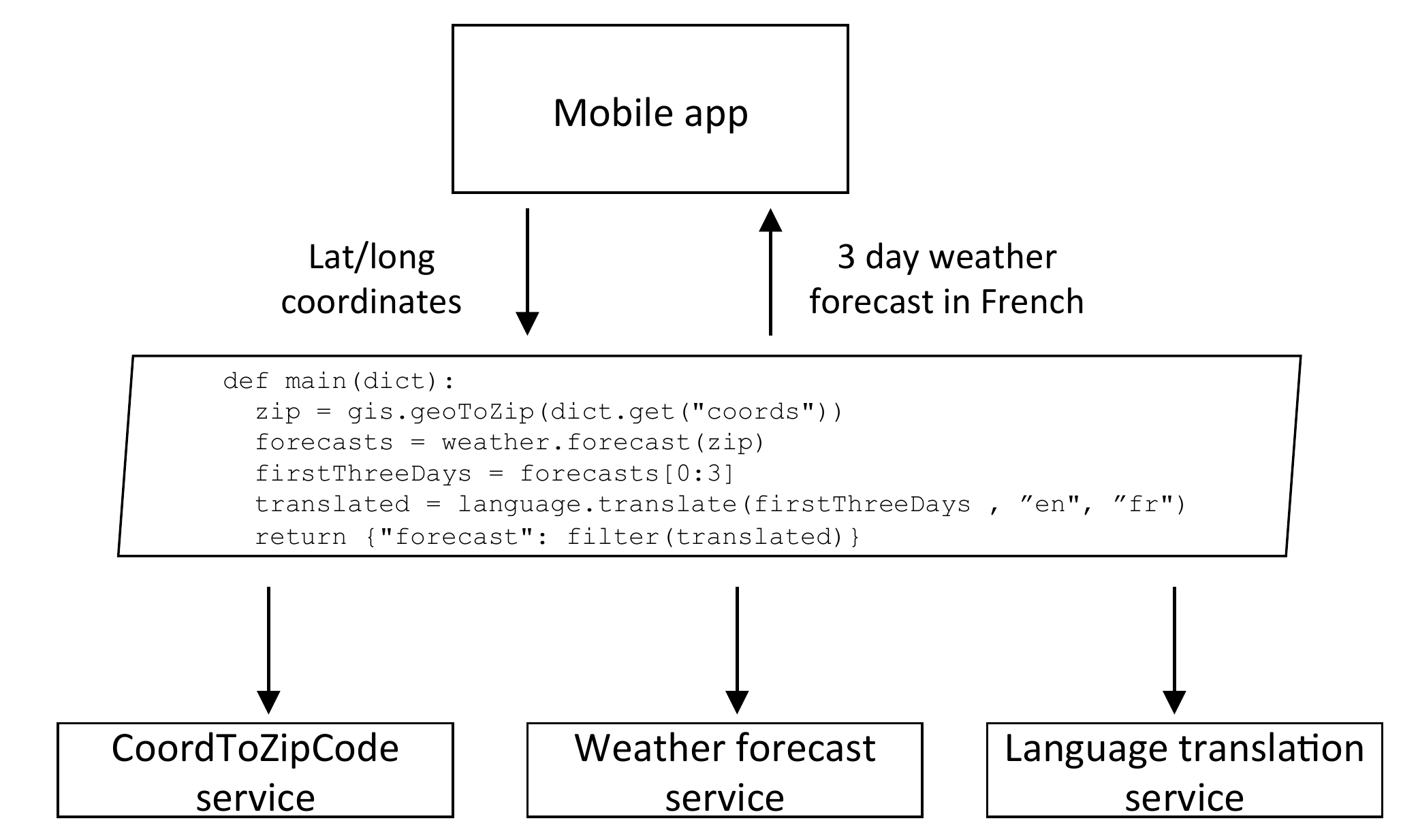}
\caption{Offloading API calls and glue logic from mobile app to
backend}
\label{fig:UseCase-Weather}
\end{figure}

Another class of applications involves the composition of a number of
APIs. In this case, the application logic consists of data filtering
and transformation. For example, a mobile app may invoke geo-location,
weather, and language translation APIs to render the weather forecast
for a user's current location. The glue code to invoke these APIs can
be written in a short serverless function, as illustrated by the
Python function in Figure~\ref{fig:UseCase-Weather}. In this way, the
mobile app avoids the cost of invoking the multiple APIs over a
potentially resource constrained mobile network connection, and
offloads the filtering and aggregation logic to the backend.

\subsection{API Aggregation to Reduce API Calls}
\begin{figure}[h]
\includegraphics[width=\textwidth]{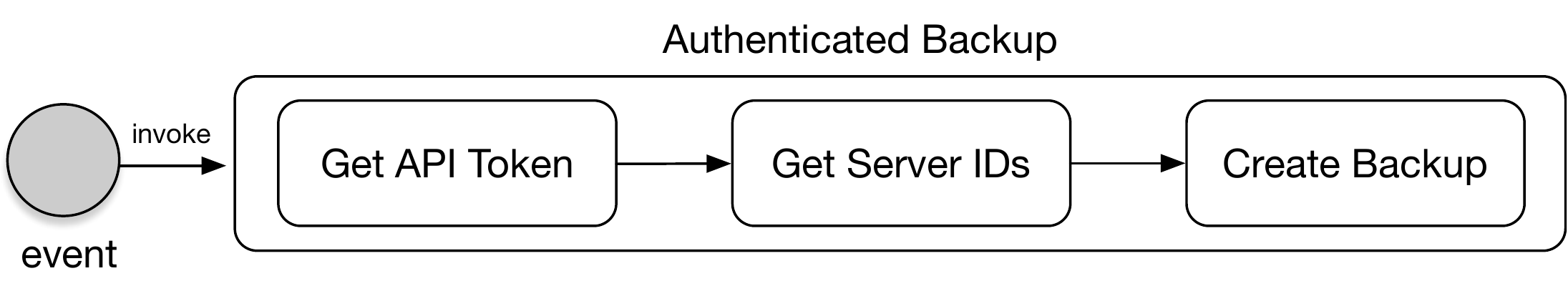}
\caption{Reducing the number of API calls required for a mobile client}
\label{fig:UseCase-Backup}
\end{figure}

API aggregation can work not only as a composition mechanism, but also as a means to simplify the client-side code that interacts with the aggregated call.  For example, consider a mobile application that allows you to administer an Open Stack instance.  API calls in Open Stack \cite{openstack} require the client to first obtain an API token, resolve the URL of the service you need to talk to, then invoke the required API call on that URL with the API token. Ideally, a mobile app would save energy by minimizing the number of required calls needed to issue a command an Open Stack instance.  Figure ~\ref{fig:UseCase-Backup} illustrates an alternative approach where three functions implement the aforementioned flow to allow authenticated backups in an Open Stack instance.  The mobile client now makes a single call to invoke this aggregate function. The flow itself appears as a single API call.  Note that authorization to invoke this call can be handled by an external authorization service, e.g. an API gateway.

\subsection{Flow control for Issue Tracking}
\begin{figure}[h]
\includegraphics[width=\textwidth]{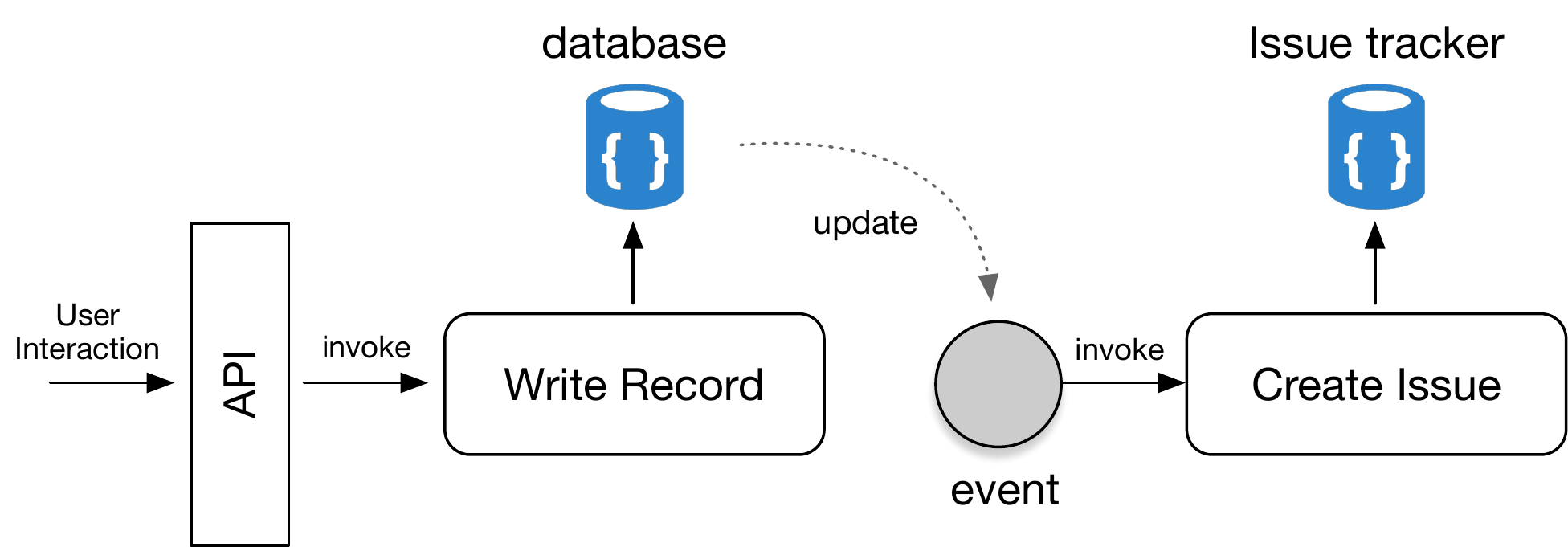}
\caption{Batched invocation for issue tracking}
\label{fig:UseCase-Issues}
\end{figure}

Serverless function composition can be used to control the flow of data between two services. For example, imagine an application that allows users to submit feedback to the app developers in the form of annotated screenshots and text. In figure ~\ref{fig:UseCase-Issues}, the application submits this data to a backend consisting of a scalable database and an on-premise issue tracking system.  The latter is mainly used by the development team and is not designed to accept high volume traffic. On the other hand, the former is capable of responding to high volume traffic.  We design our system to stage all feedback records in the database using a serverless function which eliminates the need to standup a separate server to handle feedback requests but still allows us a level of indirection between the application and the backend database. Once we collect a sufficient number of updates, we can batch them together into a single update, which invokes a function to submit issues to the issue tracker in a controlled manner. This flow would work for a scalable database system ~\cite{cloudant} and an issue tracker system that accepts batched inputs \cite{jira}.

\subsection{Discussion}

The workload and its relationship to cost can help determine if serverless is appropriate. Infrequent but bursty workloads may be better served by serverless, which provides horizontal scaling without the need for dedicated infrastructure that charges for idle time. For more steady workloads, the frequency at which a function is executed will influence how economical it can be for caching to occur which allows far faster execution on warm containers than executing from a cold container. These performance characteristics can help guide the developer when considering serverless.

Interestingly, the cost considerations may affect how a serverless
application is structured. For example, an I/O bound serverless
function can be decomposed into multiple compute bound ones. This may
be more complex to develop and debug, but cheaper to operate.


\section{Challenges and open problems} \label{sec:challenges}

We will list challenges starting with those that are already known
based on our experience of using serverless services
and then describe open problems.


\subsection{System level challenges}

Here is a list of challenges at the systems level.

\begin{itemize}
    \item \emph{Cost}: Cost is a fundamental challenge. This
        includes minimizing the resource usage of a serverless
        function, both when it is executing and when idle.
        Another aspect is the pricing model, including how it
        compares to other cloud computing approaches. For
        example, serverless functions are currently most
        economical for CPU-bound computations, whereas I/O
        bound functions may be cheaper on dedicated VMs or
        containers.

    \item \emph{Cold start}: A key differentiator of serverless is
        the ability to scale to zero, or not charging
        customers for idle time. Scaling to zero, however,
        leads to the problem of cold starts, and paying the
        penalty of getting serverless code ready to run. Techniques
        to minimize the cold start problem while still scaling
        to zero are critical.

    \item \emph{Resource limits}: Resource limits are needed to
        ensure that the platform can handle load spikes, and
        manage attacks. Enforceable resource limits on a
        serverless function include memory, execution time,
        bandwidth, and CPU usage. In additional, there are
        aggregate resource limits that can be applied across a
        number of functions or across the entire platform.

    \item \emph{Security}: Strong isolation of functions is
        critical since functions from many users are running
        on a shared platform.

    \item \emph{Scaling}: The platform must ensure the scalability
        and elasticity of users' functions. This includes
        proactively provisioning resources in response to
        load, and in anticipation of future load. This is a
        more challenging problem in serverless because these
        predictions and provisioning decisions must be made
        with little or no application-level knowledge. For
        example, the system can use request queue lengths as
        an indication of the load, but is blind to the nature
        of these requests.

    \item \emph{Hybrid cloud}: As serverless is gaining popularity
        there may be more than one serverless platform and
        multiple serverless services that need to work together.
        It is unlikely one platform will have all functionality
        and work for all use cases.

    \item \emph{Legacy systems}: It should be easy
        to access older cloud and non-cloud systems from serverless code
        running in serverless platforms.
\end{itemize}

\subsection{Programming model and DevOps challenges}

\begin{itemize}
    \item \emph{Tools}: Traditional tools that assumed access to servers
        to be able to monitor and debug applications aren't applicable in
        serverless architectures, and new approaches are needed.

    \item \emph{Deployment}: Developers should be able to use declarative approaches to
        control what is deployed and tools to support it.

    \item \emph{Monitoring and debugging}: As developers no longer have servers
        that they can access,  serverless services and tools need
        to focus on developer productivity.
        As serverless functions are running for shorter amounts of time
        there will be many orders of magnitude more of them running
        making it harder to identify problems and bottlenecks.
        When the functions finish the only trace of their execution is what
        serverless platforms monitoring recorded.

    \item \emph{IDEs}: Higher level developer capabilities, such
        as refactoring functions (e.g., splitting and merging
        functions), reverting to an older version, etc. will be needed
        and should be fully integrated with serverless platforms.

    \item \emph{Composability}: This includes being able to call
        one function from another, creating functions that
        call and coordinate a number of other functions, and
        higher level constructs such as parallel executions
        and graphs. Tools will be needed to facilitate creation
        of compositions and their maintenance.

    \item \emph{Long running}: Currently serverless functions are
        often limited in their execution time. There are
        scenarios that require long running (if intermittent)
        logic. Programming models and tools may decompose
        long running tasks into smaller units and provide
        necessary context to track them as one long running unit of work.

    \item \emph{State}: Real applications often require state, and
        it's not clear how to manage state in stateless
        serverless functions - programing models, tools, libraries etc.
        will need to provide necessary levels of abstraction.

    \item \emph{Concurrency}: Express concurrency semantics, such
        as atomicity (function executions need to be
        serialized), etc.

    \item \emph{Recovery semantics}: Includes exactly once, at
        most once, and at least once semantics.

    \item \emph{Code granularity}: Currently, serverless platforms
        encapsulate code at the granularity of functions.
        It's an open question whether coarser or finer grained
        modules would be useful.

\end{itemize}

\subsection{Open Research Problems}

Now we will describe a number of open problems. We frame them as
questions to emphasize that they are largely unexplored research
areas.

\textbf{What are the boundaries of serverless?}
A fundamental question about serverless computing is of
boundaries: is it restricted to FaaS or is broader in scope? How does it
relate to other models such as SaaS and MBaaS?

\begin{figure}[h]
\includegraphics[width=\textwidth]{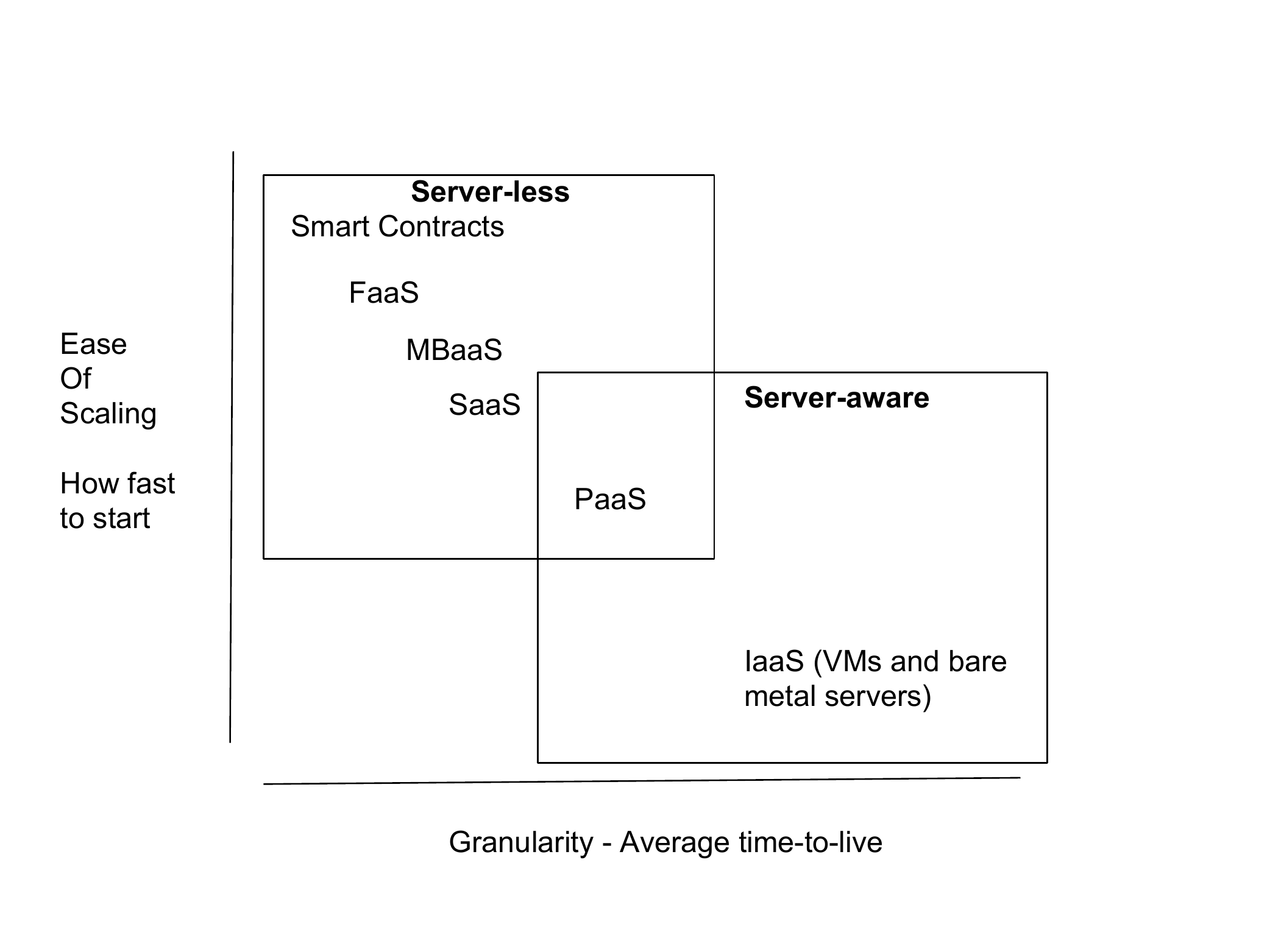}
\caption{The figure is showing relation between time-to-live (x-axis) and ease-of-scaling (y-axis). Server-aware compute (bare metal, VMs, IaaS) has long time to live and take longer to scale (time to provision new resources); server-less compute (FaaS, MBaaS, PaaS, SaaS) is optimized to work on multiple servers and hide server details.}
\label{fig:granularity}
\end{figure}

As serverless is gaining popularity the boundaries between different types of "as-a-Service" may be disappearing (see Figure \ref{fig:granularity}). One could imagine that developers not only write code but also declare how they want the code to run - as FaaS or MBaaS or PaaS - and can change as needs change.
In the future the main distinction may be between caring about server
(server-aware) and not caring about server details (server-less).
PaaS is in the middle; it makes it very easy to deploy code but
developers still need to know about servers and be aware of scaling
strategies, such as how many instances to run.


Can different cloud computing service models be mixed? Can there be more choices for how much memory and CPU can be used by serverless functions? Does serverless need to have IaaS-like based pricing? What about spot and dynamic pricing with dynamically changing granularity?

\textbf{Is tooling for serverless fundamentally different from
existing solutions?}
As the granularity of serverless is much smaller than traditional
server based tools we may need new tools
to deal well with more numerous but much shorter
living artifacts. How can we make sure that the important "information
needle" is not lost in haystack?  Monitoring and debugging serverless
applications will be much more challenging as there are no servers
directly accessible to see what went wrong. Instead, serverless
platforms need to gather all data when code is running and make it
available later. Similarly debugging is much different if instead of
having one artifact (a micro-service or traditional monolithic app)
developers need to deal with a myriad of smaller pieces of code. New
approaches may be needed to virtually assemble serverless pieces into
larger units that are easier to understand and to reason about.

\textbf{Can legacy code be made to run serverless?}
The amount of existing (``legacy") code that must continue running is
much larger than the new code created specifically to run in
serverless environments. The economical value of existing code
represents a huge investment of countless hours of developers coding
and fixing software. Therefore, one of the most important problems may
be to what degree existing legacy code can be automatically or
semi-automatically decomposed into smaller-granularity
pieces to take advantage of these new economics.

\textbf{Is serverless fundamentally stateless?}
As current serverless platforms are stateless will there be stateful serverless services in future?
Will there be simple ways to deal with state? More than that: is serverless fundamentally stateless? Can there be serverless services that have stateful support built-in with different degrees of quality-of-service?

\textbf{Will there be patterns for building serverless solutions?}
How do we combine low granularity “basic” building blocks of serverless into bigger solutions?
How are we going to decompose apps into functions so that they optimize resource usage? For example how do we identify CPU-bound parts of applications built to run in serverless services?
Can we use well-defined patterns for composing functions and external APIs?
What should be done on the server vs. client (e.g., are thicker clients more appropriate here)?
Are there lessons learned that can be applied from OOP design patterns, Enterprise Integration Patterns, etc.?

\textbf{Does serverless extend beyond traditional cloud platforms?}
Serverless may need to support scenarios where code is executed outside of a traditionally defined data center. This may include efforts where cloud is extended to include IoT,
mobile devices, web browsers, and other computing at the edge. For example ``fog"
computing \cite{fog-computing} has the goal of creating a system-level
horizontal architecture that distributes resources and services of
computing, storage, control and networking anywhere along the
continuum from Cloud to IoT. The code running in the ``fog" and
outside the Cloud may not just be embedded but virtualized to allow
movement between devices and cloud. That may lead to specific
requirements that redefine cost. For example, energy usage may be more
important than speed.

Another example is running code that executes ``smart contracts" orchestrating transactions in BlockChain. The code that defines the contract may be deployed and running on a network of Hyperledger fabric peer nodes \cite{chaincode}, or in Ethereum Virtual Machines \cite{ethereum} on any node of an Ethereum peer-to-peer network. As the system is decentralized there is no Ethereum service or servers to run serverless code. Instead, to incentivize Ethereum users to run smart contracts they get paid for the “gas” consumed by the code, similar to fuel cost for an automobile but applied to computing.

\section{Conclusions}

In this chapter we explored the genesis and history of serverless
computing in detail. It is an evolution of the trend towards higher
levels of abstractions in cloud programming models, and currently
exemplified by the Function-as-a-Service (FaaS) model where developers
write small stateless code snippets and allow the platform to manage
the complexities of scalably executing the function in a
fault-tolerant manner.

This seemingly restrictive model nevertheless lends itself well to a
number of common distributed application patterns, including
compute-intensive event processing pipelines. Most of the large cloud
computing vendors have released their own serverless platforms, and
there is a tremendous amount of investment and attention around this
space in industry.

Unfortunately, there has not been a corresponding degree of interest
in the research community. We feel strongly that there are a wide
variety of technically challenging and intellectually deep problems in
this space, ranging from infrastructure issues such as optimizations
to the cold start problem to the design of a composable programming
model. There are even philosophical questions such as the fundamental
nature of state in a distributed application. Many of the open
problems identified in this chapter are real problems faced by
practitioners of serverless computing today and solutions have the
potential for significant impact.



We leave the reader with some ostensibly simple questions that we hope
will help stimulate their interest in this area. Why is serverless
computing important?  Will it change the economy of computing? As
developers take advantage of smaller granularities of computational
units and pay only for what is actually used will that change how
developers think about building solutions? In what ways will
serverless extend the original intent of cloud computing of making
hardware fungible and shifting cost of computing from capital to
operational expenses?

The serverless paradigm may eventually lead to new kinds of
programming models, languages, and platform architectures and that is
certainly an exciting area for the research community to participate
in and contribute to.


%
\bibliographystyle{spmpsci}
\bibliography{references}
%

\end{document}